\def\pp{{\mathchoice
          {
              \kern 1pt%
              \raise 1pt
              \vbox{\hrule width5pt height0.4pt depth0pt
                    \kern -2pt
                    \hbox{\kern 2.3pt
                          \vrule width0.4pt height6pt depth0pt
                          }
                    \kern -2pt
                    \hrule width5pt height0.4pt depth0pt}%
                    \kern 1pt
           }
            {
              \kern 1pt%
              \raise 1pt
              \vbox{\hrule width4.3pt height0.4pt depth0pt
                    \kern -1.8pt
                    \hbox{\kern 1.95pt
                          \vrule width0.4pt height5.4pt depth0pt
                          }
                    \kern -1.8pt
                    \hrule width4.3pt height0.4pt depth0pt}%
                    \kern 1pt
            }
            {
              \kern 0.5pt%
              \raise 1pt
              \vbox{\hrule width4.0pt height0.3pt depth0pt
                    \kern -1.9pt  
                    \hbox{\kern 1.85pt
                          \vrule width0.3pt height5.7pt depth0pt
                          }
                    \kern -1.9pt
                    \hrule width4.0pt height0.3pt depth0pt}%
                    \kern 0.5pt
            }
            {
              \kern 0.5pt%
              \raise 1pt
              \vbox{\hrule width3.6pt height0.3pt depth0pt
                    \kern -1.5pt
                    \hbox{\kern 1.65pt
                          \vrule width0.3pt height4.5pt depth0pt
                          }
                    \kern -1.5pt
                    \hrule width3.6pt height0.3pt depth0pt}%
                    \kern 0.5pt
            }
        }}

\def\mm{{\mathchoice
                       {
                             \kern 1pt
               \raise 1pt    \vbox{\hrule width5pt height0.4pt depth0pt
                                  \kern 2pt
                                  \hrule width5pt height0.4pt depth0pt}
                             \kern 1pt}
                       {
                            \kern 1pt
               \raise 1pt \vbox{\hrule width4.3pt height0.4pt depth0pt
                                  \kern 1.8pt
                                  \hrule width4.3pt height0.4pt depth0pt}
                             \kern 1pt}
                       {
                            \kern 0.5pt
               \raise 1pt
                            \vbox{\hrule width4.0pt height0.3pt depth0pt
                                  \kern 1.9pt
                                  \hrule width4.0pt height0.3pt depth0pt}
                            \kern 1pt}
                       {
                           \kern 0.5pt
             \raise 1pt  \vbox{\hrule width3.6pt height0.3pt depth0pt
                                  \kern 1.5pt
                                  \hrule width3.6pt height0.3pt depth0pt}
                           \kern 0.5pt}
                       }}

\documentstyle[12pt]{article}

\skewchar\fivmi='177
\skewchar\sixmi='177
\skewchar\sevmi='177
\skewchar\egtmi='177
\skewchar\ninmi='177
\skewchar\tenmi='177
\skewchar\elvmi='177
\skewchar\twlmi='177
\skewchar\frtnmi='177
\skewchar\svtnmi='177
\skewchar\twtymi='177
\def\@magscale#1{ scaled \magstep #1}
\skewchar\fivsy='60
\skewchar\sixsy='60
\skewchar\sevsy='60
\skewchar\egtsy='60
\skewchar\ninsy='60
\skewchar\tensy='60
\skewchar\elvsy='60
\skewchar\twlsy='60
\skewchar\frtnsy='60
\skewchar\svtnsy='60
\skewchar\twtysy='60

\catcode`@=11
\def\un#1{\relax\ifmmode\@@underline#1\else
        $\@@underline{\hbox{#1}}$\relax\fi}
\catcode`@=12

\def\a{\alpha}
\def\b{\beta}

\def\d{\delta}

\def\g{\gamma}

\def\q{\theta}

\def\s{\sigma}

\def\z{\zeta}

\def\P{\Pi}

\def\dslash{\not{\hbox{\kern-2pt $\partial$}}}
\def\Dslash{\not{\hbox{\kern-4pt $D$}}}
\def\pslash{\not{\hbox{\kern-2.3pt $p$}}}
 \newtoks\slashfraction
 \slashfraction={.13}
 \def\slash#1{\setbox0\hbox{$ #1 $}
 \setbox0\hbox to \the\slashfraction\wd0{\hss \box0}/\box0 }
 
\font\ro=cmsy10                          
\def\kcr{{\hbox{\ro \char'170}}}                
\def\ktl{{\hbox{\ro \char'170}}}        
\def\ktr{{\hbox{\ro \char'170}}}        
\def\kbl{{\hbox{\ro \char'170}}}        
\def\kbr{{\hbox{\ro \char'170}}}        

\def\plpl{\raise-2pt\hbox{$\raise3pt\hbox{$_+$}\hskip-6.67pt\raise0.0pt
\hbox{$^+$}\hskip 0.01pt$}}
\def\mimi{\raise-2pt\hbox{$\raise3pt\hbox{$_-$}\hskip-6.67pt\raise0.0pt
\hbox{$^-$}\hskip 0.01pt$}} 

\def\bo{{\raise.15ex\hbox{\large$\Box$}}}               
\def\pa{\partial}                                       
\def\TH{{\raise.2ex\hbox{$\displaystyle \bigodot$}\mskip-4.7mu \llap H \;}}
\def\face{{\raise.2ex\hbox{$\displaystyle \bigodot$}\mskip-2.2mu \llap {$\ddot
        \smile$}}}                                      
\def\dg{\sp\dagger}                                     

\def\pp{{\mathchoice
          {
              \kern 1pt%
              \raise 1pt
              \vbox{\hrule width5pt height0.4pt depth0pt
                    \kern -2pt
                    \hbox{\kern 2.3pt
                          \vrule width0.4pt height6pt depth0pt
                          }
                    \kern -2pt
                    \hrule width5pt height0.4pt depth0pt}%
                    \kern 1pt
           }
            {
              \kern 1pt%
              \raise 1pt
              \vbox{\hrule width4.3pt height0.4pt depth0pt
                    \kern -1.8pt
                    \hbox{\kern 1.95pt
                          \vrule width0.4pt height5.4pt depth0pt
                          }
                    \kern -1.8pt
                    \hrule width4.3pt height0.4pt depth0pt}%
                    \kern 1pt
            }
            {
              \kern 0.5pt%
              \raise 1pt
              \vbox{\hrule width4.0pt height0.3pt depth0pt
                    \kern -1.9pt  
                    \hbox{\kern 1.85pt
                          \vrule width0.3pt height5.7pt depth0pt
                          }
                    \kern -1.9pt
                    \hrule width4.0pt height0.3pt depth0pt}%
                    \kern 0.5pt
            }
            {
              \kern 0.5pt%
              \raise 1pt
              \vbox{\hrule width3.6pt height0.3pt depth0pt
                    \kern -1.5pt
                    \hbox{\kern 1.65pt
                          \vrule width0.3pt height4.5pt depth0pt
                          }
                    \kern -1.5pt
                    \hrule width3.6pt height0.3pt depth0pt}%
                    \kern 0.5pt
            }
        }}

\def\sp#1{{}^{#1}}                              
\def\Tilde#1{\widetilde{#1}}                    
\def\Hat#1{\widehat{#1}}                        
\def\Bar#1{\overline{#1}}                       
\def\leftrightarrowfill{$\mathsurround=0pt \mathord\leftarrow \mkern-6mu
        \cleaders\hbox{$\mkern-2mu \mathord- \mkern-2mu$}\hfill
        \mkern-6mu \mathord\rightarrow$}
\def\dvec#1{\vbox{\ialign{##\crcr
        \leftrightarrowfill\crcr\noalign{\kern-1pt\nointerlineskip}
        $\hfil\displaystyle{#1}\hfil$\crcr}}}           

\def\frac#1#2{{\textstyle{#1\over\vphantom2\smash{\raise.20ex
        \hbox{$\scriptstyle{#2}$}}}}}                   
\def\sfrac#1#2{{\vphantom1\smash{\lower.5ex\hbox{\small$#1$}}\over
        \vphantom1\smash{\raise.4ex\hbox{\small$#2$}}}} 
\def\bfrac#1#2{{\vphantom1\smash{\lower.5ex\hbox{$#1$}}\over
        \vphantom1\smash{\raise.3ex\hbox{$#2$}}}}       
\def\afrac#1#2{{\vphantom1\smash{\lower.5ex\hbox{$#1$}}\over#2}}    
\def\partder#1#2{{\partial #1\over\partial #2}}   

\newskip\humongous \humongous=0pt plus 1000pt minus 1000pt
\def\caja{\mathsurround=0pt}
\def\eqalign#1{\,\vcenter{\openup2\jot \caja
        \ialign{\strut \hfil$\displaystyle{##}$&$
        \displaystyle{{}##}$\hfil\crcr#1\crcr}}\,}
\newif\ifdtup

\def\ref#1{$\sp{#1)}$}

\parskip=\medskipamount                 
\thicklines                         

\def\oldheadpic{                                
        \setlength{\unitlength}{.4mm}
        \thinlines
        \par
        \begin{picture}(349,16)
        \put(325,16){\line(1,0){4}}
        \put(330,16){\line(1,0){4}}
        \put(340,16){\line(1,0){4}}
        \put(335,0){\line(1,0){4}}
        \put(340,0){\line(1,0){4}}
        \put(345,0){\line(1,0){4}}
        \put(329,0){\line(0,1){16}}
        \put(330,0){\line(0,1){16}}
        \put(339,0){\line(0,1){16}}
        \put(340,0){\line(0,1){16}}
        \put(344,0){\line(0,1){16}}
        \put(345,0){\line(0,1){16}}
        \put(329,16){\oval(8,32)[bl]}
        \put(330,16){\oval(8,32)[br]}
        \put(339,0){\oval(8,32)[tl]}
        \put(345,0){\oval(8,32)[tr]}
        \end{picture}
        \par
        \thicklines
        \vskip.2in}
\def\oldtitle#1#2#3#4{\oldheadpic\begin{center}\vglue.5in{\large\bf #1}\\[.6in]
        {#2}\\[.1in] {\it Department of Physics and Astronomy}\\
        {\it University of Maryland, College Park, MD 20742}\\[.6in]
        Physics Publication \#{#3}\\ {#4}\\[1.5in] {\bf ABSTRACT}\\[.1in]
        \end{center} \begin{quotation}}                 
\def\oldTitle#1#2#3#4#5#6#7{\oldheadpic\begin{center} \vglue .4in
        {\large\bf #1}\\[.4in]
        {#2}\\[.1in] {\it Department of Physics and Astronomy}\\
        {\it University of Maryland, College Park, MD 20742}\\[.1in]
        {#3}\\[.1in] {\it {#4}}\\ {\it {#5}}\\[.4in]
        Physics Publication \#{#6}\\ {#7}\\[.5in] {\bf ABSTRACT}\\[.1in]
        \end{center} \begin{quotation}}                 
\def\border{                                            
        \setlength{\unitlength}{1mm}
        \newcount\xco
        \newcount\yco
        \xco=-21
        \yco=12
        \begin{picture}(140,0)
        \put(\xco,\yco){$\ktl$}
        \advance\yco by-1
        {\loop
        \put(\xco,\yco){$\kcr$}
        \advance\yco by-2
        \ifnum\yco>-240
        \repeat
        \put(\xco,\yco){$\kbl$}}
        \xco=158
        \yco=12
        \put(\xco,\yco){$\ktr$}
        \advance\yco by-1
        {\loop
        \put(\xco,\yco){$\kcr$}
        \advance\yco by-2
        \ifnum\yco>-240
        \repeat
        \put(\xco,\yco){$\kbr$}}
        \put(-20,13){\tiny University of Maryland Elementary Particle
Physics University of Maryland Elementary Particle Physics University of
Maryland Elementary Particle Physics}
        \put(-20,-241.5){\tiny University of Maryland Elementary
Particle Physics University of Maryland Elementary Particle Physics
University of Maryland Elementary Particle Physics}
        \end{picture}
        \par\vskip-8mm}
\def\bordero{                                           
        \setlength{\unitlength}{1mm}
        \newcount\xco
        \newcount\yco
        \xco=-31
        \yco=12
        \begin{picture}(140,0)
        \put(\xco,\yco){$\ktl$}
        \advance\yco by-1
        {\loop
        \put(\xco,\yco){$\kclr}
        \advance\yco by-2
        \ifnum\yco>-240
        \repeat
        \put(\xco,\yco){$\kbl$}}
        \xco=151
        \yco=12
        \put(\xco,\yco){$\ktr$}
        \advance\yco by-1
        {\loop
        \put(\xco,\yco){$\kcr$}
        \advance\yco by-2
        \ifnum\yco>-240
        \repeat
        \put(\xco,\yco){$\kbr$}}
        \put(-20,12){\ooo bacdefghidfghghdhededbihdgdfdfhhdheidhdhebaaahjhhdahba

hgdedge
   hgfdiehhgdigicba}
        \put(-20,-241.5){\ooo ababaighefdbfghgeahgdfgafagihdidihiidhiagfedhadbfd

ecdcdfa
   gdcbhaddhbgfchbgfdacfediacbabab}
        \end{picture}
        \par\vskip-8mm}
\def\headpic{                                           
        \indent
        \setlength{\unitlength}{.4mm}
        \thinlines
        \par
        \begin{picture}(29,16)
        \put(165,16){\line(1,0){4}}
        \put(170,16){\line(1,0){4}}
        \put(180,16){\line(1,0){4}}
        \put(175,0){\line(1,0){4}}
        \put(180,0){\line(1,0){4}}
        \put(185,0){\line(1,0){4}}
        \put(169,0){\line(0,1){16}}
        \put(170,0){\line(0,1){16}}
        \put(179,0){\line(0,1){16}}
        \put(180,0){\line(0,1){16}}
        \put(184,0){\line(0,1){16}}
        \put(185,0){\line(0,1){16}}
        \put(169,16){\oval(8,32)[bl]}
        \put(170,16){\oval(8,32)[br]}
        \put(179,0){\oval(8,32)[tl]}
        \put(185,0){\oval(8,32)[tr]}
        \end{picture}
        \par\vskip-6.5mm
        \thicklines}
\def\title#1#2#3#4{\border\headpic {\hbox to\hsize{#4 \hfill UMDEPP #3}}\par
        \begin{center} \vglue .5in {\large\bf #1}\\[.6in]
        {#2}\\[.1in] {\it Department of Physics and Astronomy}\\
        {\it University of Maryland, College Park, MD 20742}\\[1.5in]
        {\bf ABSTRACT}\\[.1in] \end{center} \begin{quotation}}  
\def\Title#1#2#3#4#5#6#7{\border\headpic
        {\hbox to\hsize{#7 \hfill UMDEPP #6}}\par
        \begin{center} \vglue .4in {\large\bf #1}\\[.4in]
        {#2}\\[.1in] {\it Department of Physics and Astronomy}\\
        {\it University of Maryland, College Park, MD 20742}\\[.1in]
        {#3}\\[.1in] {\it {#4}}\\ {\it {#5}}\\[.5in] {\bf ABSTRACT}\\[.1in]
        \end{center} \begin{quotation}}                 
\def\endtitle{\end{quotation}\newpage}                  

\def\ad{{\kern0.5pt
                   \alpha \kern-5.05pt \raise5.8pt\hbox{$\textstyle.$}\kern
0.5pt}}

\def\bd{{\kern0.5pt
                   \beta \kern-5.05pt \raise5.8pt\hbox{$\textstyle.$}\kern
0.5pt}}

\def\qd{{\kern0.5pt
                   q \kern-5.05pt \raise5.8pt\hbox{$\textstyle.$}\kern
0.5pt}}
\def\Dot#1{{\kern0.5pt
     {#1} \kern-5.05pt \raise5.8pt\hbox{$\textstyle.$}\kern
0.5pt}}

\begin{document}

\def\gfrac#1#2{\frac {\scriptstyle{#1}}
        {\mbox{\raisebox{-.6ex}{$\scriptstyle{#2}$}}}}
\def\gg{{\hbox{\sc g}}}
\border\headpic {\hbox to\hsize{January 1997 \hfill {UMDEPP 97-74}}}
\par
\setlength{\oddsidemargin}{0.3in}
\setlength{\evensidemargin}{-0.3in}
\begin{center}
\vglue 1in
{\large\bf What If Dirac Pionini Existed\\
in a Purely Chiral Superfield Formulation?\footnote {Supported 
in part by National Science Foundation Grant PHY-91-19746 
\newline ${~~~~~}$ and by NATO Grant CRG-93-0789}  }
\\[.72in]

S. James Gates, Jr.\footnote{gates@umdhep.umd.edu} and Lubna 
Rana\footnote{lubna@umdhep.umd.edu}
\\[.02in]
{\it Department of Physics\\ 
University of Maryland\\ 
College Park, MD 20742-4111  USA}\\[2in]

{\bf ABSTRACT}\\[.002in]
\end{center}
\begin{quotation}
{By an explicit construction, it is shown that the geometry of 
the $SU(3)$ pion multiplet with respect to the group manifold 
$SU_L(3) \otimes SU_R(3)$ may be deformed to admit a second 
pseudoscalar multiplet that is analogous to the $Z^0$ in 
unified theories of the electroweak interaction. This observation 
is found to play a key role in the construction of the N = 1 
supersymmetric models with pions and Dirac-like spin-1/2 
superpartners (`pionini'). }  

\endtitle
\section{Introduction} 

~~~~In a recent series of papers \cite{SJG,JOMMM}, an
exploration of the possibility to construct 4D, N = 1 superfield 
actions that generalize the phenomenological approach of Chiral 
Perturbation Theory \cite{chipt} has been carried out.  With the 
work of \cite{JOMMM}, the initial phase of the construction of 
super-Chiral Perturbation Theory (or ``S-Chiral Perturbation Theory'') 
was completed.  Of course, there is no sign that the low-energy 
physics of pions is supersymmetric, but in a hypothetical world of 
unbroken supersymmetry, there is no impediment to the study of such 
a model.  

In such a hypothetical world, accompanying the pion there would 
exist a set of spin-1/2 particles (hereafter to be called ``pionini'').  
Carrying these consideration one step further, we are confronted 
with a choice of whether the pionini should be Dirac, Majorana or 
Weyl particles.  This makes a big difference in the attempt to 
construct models of the QCD effective action in this hypothetical 
world.  In the former case, it would be required that there are 
{\it {two}} $SU(3)$-octets of 4D, N = 1 superfields to construct 
the supermultiplet containing the pionini.  In the latter two cases, 
only one $SU(3)$-octet of 4D, N = 1 superfields is required.

For our purposes, let us assume that the hypothetical pionini are 
Dirac particles.  (One reason for this assumption is that most of 
the existing literature on 4D, N = 1 supersymmetric QCD-like effective 
actions\footnote{For a nice review of a large part of the literature 
on this topic, see ref. \cite{1}. A discussion to \newline ${~\,~~~}$
consider more recent developments can be found in the work of ref. 
\cite{2}.} is based on the opposite assumption.  Thus that case 
is well studied already.)  On the other hand, it is only with our 
recent work \cite{SJG} that models which describe Dirac pionini have 
entered the literature. However, the class of models that we have 
described so far utilizes {\it {both}} chiral (C), i.e. Wess-Zumino 
\cite{WZ}, and nonminimal (N)  \cite{GS}, i.e. complex linear, 
superfields to describe the two chiral components of a Dirac spinor.  
Since such a description uses both types of multiplets, we call 
these CNM (chiral-nonminimal multiplet) models.  On the other hand, 
almost all of the literature on phenomenological applications of 
supersymmetry \cite{moha} uses pairs of chiral superfields to describe 
Dirac spinors.  We are thus motivated to ask, ``Whether a 4D, N = 1 
supersymmetric QCD-like effective action containing Dirac pionini 
is possible to construct utilizing {\it {only}} chiral superfields?'' 
The answer to this question turns out to be affirmative as we will 
see in this work.

As a by-product of this question posed within the context of 
supersymmetrical theories, we will also learn how it is possible to 
formulate the $SU_L(3) \otimes SU_R(3)$ geometry of the pion in a 
manner that has not (at least to our knowledge) appeared in the physics 
literature previously.

\section{Ur-formulations of $SU_L(3) \otimes SU_R(3)$ Geometry of 
the Pion}

~~~~For as long as it has been known that the group manifold
$SU_L(3) \otimes SU_R(3)$ plays an important role in understanding
the dynamics of pions, it has been assumed the the interpretation
of this particle in terms of the geometry of this manifold is
unique. The pion octet usually appears as an element of the $SU(3)$
algebra in the form,
$$ \frac 1{f_{\pi}} \Pi ~\equiv~ \frac 1{ f_{\pi}} \Pi^i t_i ~=~ 
\frac 1{f_{\pi}} \left(\begin{array}{ccc}
{}~\frac{\pi^0}{\sqrt 2} ~+~ \frac{\eta}{\sqrt 6} & ~~\pi^+ & 
~~K^+ \\
{}~\pi^- & ~~-\, \frac{\pi^0}{\sqrt 2} ~+~ \frac{\eta}{\sqrt 6} & 
~~K^0\\ 
{}~K^- & ~~{\Bar K}^0 & ~~ - \eta \sqrt {\frac 23} \\
\end{array}\right) ~~~~.
\eqno(1) $$
where $t_1 , \,..., t_8$ are essentially the Gell-Mann $SU(3)$ matrices 
and $f_{\pi}$ is the pion decay constant. The $SU(3)$ group elements 
are obtained via exponentiation  $U \equiv \exp [ i \frac 1{ f_{\pi}} 
\Pi^i t_i ]$. The rigid $SU_L(3) \otimes SU_R(3)$ transformations 
that are symmetries of the effective action are defined by
$$
\Big( U \Big)' ~=~ \exp [ - i {\Tilde \a}^i t_i \,] ~ U
{}~ \exp [ i {\a}^i t_i \,] ~~~,
\eqno(2)
 $$
with independent real left and right transformation parameters 
${\Tilde \a}^i$ and $\a^i$. For infinitesimal parameters this can be 
written as a variation of the $SU(3)$ octet containing the pion fields, 
$$
\d \Pi^i ~=~ - \, i \, f_{\pi}\,  [\, 
{\Tilde \a}^j (L^{-1} )_j {}^i ~-~
{\a}^j (R^{-1} )_j {}^i ~] \equiv  \a^{(A)} \xi^i_{(A)} 
{}~~~. \eqno(3)  $$
We note that  inverse Maurer-Cartan forms $(L^{-1} )_j {}^i$ and 
$(R^{-1} )_j {}^i$ appear.  In finite form, this corresponds to 
the coordinate transformation,
$$
\Big( \Pi^i  \Big)' ~=~ K^i (\Pi ) ~ = ~ \exp [ \, \a^{(A)} 
\xi^j_{(A)} \pa_j \,]~ \Pi^i ~~~,~~~  \pa_j ~\equiv ~ \pa/ \pa 
\Pi^j ~~~.
\eqno(4) $$
Finally the $SU_L(3) \otimes SU_R(3)$ invariant action takes the
form,
$$ 
{\cal S}_{\s}(\Pi)  ~=~ - \frac {f_{\pi}^2}{2 C_2} \, \int d^4 x ~
{\rm {Tr}} [\, ( \pa^{\un a} U^{-1} \,) ~ (\pa_{\un a} U \,) \,]
{}~~~~. \eqno(5) $$
Our notational conventions have been explained in the work of
ref. \cite{JOMMM}.

Historically, the pion was introduced to be the fundamental mediator 
of the strong nuclear force much as the photon is the fundamental 
mediator of the electromagnetic force. Let us now carry this analogy 
a step (maybe too much) farther.  With the construction of the 
Glashow-Weinberg-Salam model \cite{GSW}, it was recognized that the 
photon must be accompanied by a second {\it {massive}} neutral vector 
boson, the ${\rm Z}^0$-particle.  We now wish to appeal to this 
historical precedent.  Let us imagine that the pion, like the photon, 
is actually a linear combination of two ``ur-fields.''  For the photon 
these ur-fields are the hypercharge gauge field $Y_{\un a}$ and the 
third component of isospin gauge field $W^{3}_{\un a}$.

To play the role of the ur-fields for the pion we introduce
two pseudo-scalar fields denoted by ${A}^i(x)$ and ${B}^i(x)$
$$ \eqalign{ {~~~~~~~~~}
&{A}^i(x) ~=~  \Big[ \, \Pi^i (x) cos(\g_{\rm S} ) ~+~ 
\Theta^i (x) sin(\g_{\rm S} ) \, \Big] ~~~, \cr
&{B}^i(x) ~=~ \Big[ \, - \Pi^i (x) sin(\g_{\rm S} ) ~+~ 
\Theta^i (x) cos(\g_{\rm S} ) \, \Big] ~~~.}
\eqno(6) $$
We imagine that the pseudo-scalar spin-0 fields ${\Pi}^i$ and 
${\Theta}^i$ are the ``physical states'' with the latter playing
the role of the ${\rm Z}^0$-particle. It is obvious that the
quantity $\g_S$ here plays the analogous role of the weak mixing angle, 
$\theta_W$. 

The next problem to confront is that of to what representation of 
the $SU_L(3) \otimes SU_R(3)$ symmetry should ${A}^i$ and ${B}^i$ 
belong?  Following the precedent of the Glashow-Weinberg-Salam model
wherein $Y_{\un a}$ and $W^3_{\un a}$ are assigned to distinct
representations of $SU_W(2) \otimes U_Y(1)$, we assign the pseudoscalar
 ur-fields to distinct representations. We do this by assuming
that ${A}^i$ transforms like a coordinate of the Lie algebra manifold 
but that ${B}^i$ transforms like a 1-form.  This statement implies 
that the transformation laws for ${A}^i$ are obtained by simply 
replacing ${\Pi}^i \to {A}^i$ in (2-4) above.  It is convenient 
to also make a slight re-definition in how the group elements $U$ are
obtained from the Lie algebra element $A^i$,
$$
{U} (A) ~\equiv~ \exp \Big[  {{ \, i \, A^i
t_i }\over{~ f_{\pi}\, cos (\g_{\rm S}) ~}} \Big] ~~~~~,
\eqno(7) $$
For ${B}^i$ however, the finite transformation law must take the form
$$
\Big( B^i  \Big)' ~=~ \Big( \pa K^i (A) / \pa A^j ~ \Big)  B^j ~~~.
\eqno(8) $$
or infinitesimally as
$$
\d B^i ~=~  \a^{(A)} \Big( \pa  \xi^i_{(A)}  (A) / \pa A^j ~ \Big)
B^j {}~~~. \eqno(9)  $$
Finally given these assignments of transformation properties, an action
that is invariant takes the form,
$$ 
{\cal S}_{\s}(A ,\, B) \,=\, - \Big(  \frac {f_{\pi}^2}{2 C_2} 
\Big) cos^{2} (\g_S) \int d^4 x ~ {\rm {Tr}} [\, {\cal N}_0   
 ( \pa^{\un a} U^{-1} \,) ~ (\pa_{\un a} U \,) \,+\,
{\cal N}_1 ( \pa^{\un a} {\Tilde B}\dg \,) ~ (\pa_{\un a} {\Tilde B} 
\,) \,] ~~~, \eqno(10) $$
in terms of the matrix field ${\Tilde B} \equiv i \Big( \pa U / \pa A^i
\Big) B^i$ and normalization constants ${\cal N}_0$, ${\cal N}_1$.
In order to fix these normalization constants, we note that
$$
\pa_{\un a} U ~=~ \Big(\partder{U}{A^i} \, \Big) \Big( \pa_{\un a} 
A^i \Big) ~\equiv~ \Big( \pa_i U \Big) \Big( \pa_{\un a} A^i \, 
\Big)  ~~~,~~~ \pa^{\un a} U^{-1} ~=~ \Big( \pa_i U^{-1} \Big) 
\Big( \pa^{\un a} A^i \, \Big) ~~~,
\eqno(11) $$
$${~~~~} \to {\rm {Tr}} [\, \Big(\pa^{\un a} U^{-1} \Big) \Big( 
\pa_{\un a} U \Big) \, ] ~=~ {\rm {Tr}} [\, \Big( \pa_i 
U^{-1} \Big)  \Big( \pa_j U \Big) \, ] ~ ( \pa^{\un a} A^i \, ) 
( \pa_{\un a} A^j \, )  ~~~.
\eqno(12) $$
In a similar manner, we find,
$$ \eqalign{ {~~~~} {\rm {Tr}} [\,
\Big(\pa^{\un a} {\Tilde B}\dg \Big) \Big( \pa_{\un a} {\Tilde B}
\Big)\,] ~=~ &{\rm {Tr}} [\, \Big( \pa_i U^{-1} \Big) \Big( \pa_j 
U \Big) \, ] ~ ( \pa^{\un a} B^i \, ) ( \pa_{\un a} B^j \, ) \cr
&+~ {\rm {Tr}} [\, \Big( \pa_i U^{-1} \Big) \Big( \pa_j \pa_k U 
\Big) \, ] ~ ( \pa^{\un a} B^i \, ) ( \pa_{\un a} A^j \, ) B^k \cr
&+~ {\rm {Tr}} [\, \Big( \pa_j \pa_k U^{-1} \Big) \Big( \pa_i U 
\Big) \, ] ~ ( \pa^{\un a} B^i \, ) ( \pa_{\un a} A^j \, ) B^k \cr
&+~ {\rm {Tr}} [\, \Big( \pa_i \pa_j U^{-1} \Big) \Big( \pa_k 
\pa_l U \Big) \, ]~ ( \pa^{\un a} A^i \, ) B^j ( \pa_{\un a} A^k 
\, ) B^l  ~~~, }\eqno(13) $$
where we have used the identity $U\dg = U^{-1}$.
It is also useful to note that in (11) and all subsequent equations
$$
\pa_i ~=~ \partder{~~}{A^i} ~=~ cos(\g_S ) \, \partder{~~}{{\Pi}^i}
~+~ sin(\g_S ) \, \partder{~~}{{\Theta}^i} ~~~.
\eqno(14)
$$

Now the ``physics'' of the pions described by the action in (5) is 
different from that described in (10). It is of course of interest to 
know explicitly how such differences might manifest themselves for 
example in amplitudes. In order to see this, we will first 
consider (10) in the limit where $\Theta = 0$ and calculate the 
resulting action.  This is summarized by the following equation,
$$ {\Tilde {\cal S}}_{\s}(\Pi) ~\equiv~ 
\Big[ \, \lim_{\Theta \to 0} {\cal S}_{\s}(A ,\, B) \, \Big]~=~ 
{\cal S}_{\s}(\Pi) ~+~ sin^2 (\g_S ) 
\, \Big[ \, S_1 (\Pi) ~+~ S_2 (\Pi) \,\Big] ~~~,
\eqno(15) $$
where we have chosen to set  ${\cal N}_0 = {\cal N}_1 = 1$. The 
explicit forms of the actions $S_1$ and $S_2$ are given by
$$ \eqalign{ {~~~}
S_1 (\Pi)&=~ - \Big( \frac {f_{\pi}^2}{2 C_2} \Big)  \int d^4 x ~
{\rm {Tr}} [\, \Big( \frac{\pa U\dg }{ \pa \Pi^i }  \Big)  
\Big( \frac {\pa^2 U}{ \pa \P^j  \pa \Pi^k } \Big) 
\, + \, {\rm {h.c.}} \,]~ \P^i (\, \pa^{\un a}  \P^j )
(\, \pa_{\un a}  \P^k ) ~~~, \cr
S_2 (\Pi)&= - \Big( \frac {f_{\pi}^2}{2 C_2} \Big)
\int d^4 x ~ {\rm {Tr}} [\, \Big( \frac {\pa^2 U\dg}{ \pa \P^i  
\pa \Pi^j } \Big) \, \Big( \frac {\pa^2 U }{ \pa \P^k  \pa \Pi^l } 
\Big) ] \, \P^i \, \P^k (\, \pa^{\un a}  \P^j ) (\, \pa_{\un a}  
\P^l ) ~~~. }
\eqno(16) $$
In arriving at these results we have used 
$$ \eqalign{
\lim_{\Theta \to 0} U(A) ~=~ U(\Pi) ~~~~&,~~ \lim_{\Theta \to 0} 
\Big( \frac{\pa {~~} }{ \pa A^i }  \Big) ~=~ sec (\g_S) \Big(\frac{\pa 
{~~}}{\pa {\Pi}^i }  \Big) ~~, \cr
\lim_{\Theta \to 0} A^i ~=~ \, \Pi^i \, cos( \g_S) ~~&,~~\lim_{\Theta 
\to 0} B^i ~=~ - \, \Pi^i \, sin( \g_S) ~~~.}
\eqno(17)$$
A few comments about the second limit above are in order.  Since the 
operations of taking the limit as $\Theta \to 0$ and $\partder{~~}
{{\Theta}^i}$ do not commute, we must take some care.  Due to the 
functional form of $U(A)$, the following operator equation is valid
when acting on it or any of its derivatives w.r.t. $A^i$,
$$
\partder{~~}{{\Theta}^i} ~=~ tan(\g_S) \, \partder{~~}{{\Pi}^i} ~~~.
\eqno(18)
$$
When this is inserted into (14) we arrive at the second limit
stated in (17).

It is easily seen that the differences in contributions to pure pion 
amplitudes calculated using the two distinct actions ((5) vs.~(10))
will be at least of order $sin^2(\g_S)$.  For $\g_S \sim \frac1{10}$ 
this corresponds to a suppression by a factor of one hundred between 
the differences of amplitudes computed from the two different actions. 
In principle at least (although we are pessimistic about the 
practicality of this), precision pion physics experiments should place 
upper limits on this parameter.  Additional suppression is implied by 
the reciprocal powers of $f_{\pi}$ that are implicitly contained 
in calculating $\Big( \pa U / \pa A^i \Big)$ and $\Big( \pa^2 U 
/ \pa A^i  \pa A^j \Big)$.

The process of taking the limit at $\Theta \to 0$ has one other
important consequence.  Although the two terms in (10) are separately
$SU_L(3) \otimes SU_R(3)$ invariant, after taking the limit, the
resultant action of (15) breaks this symmetry by terms of 
order $sin^2 (\g_S)$.  One way to see that the symmetry is broken
is to note that the variation of $\Theta^i$ is given by
$$
\d \Theta^i ~=~ - i f_{\pi} cos(\g_S) \, \a^{(A)} \Big\{ ~ 
sin(\g_S)  \, \xi^i_{(A)}  ~+~ cos(\g_S)  \,\, B^k (\pa_k 
\xi^i_{(A)} )  ~ \Big\} ~~~,
\eqno(19)
$$
and if we take the limit of this equations as $\Theta \to 0$ then 
there arises the constraint $\g_S = \frac {\pi}2 n$ where $n$ is any
integer. In other words, no mixing in the initial model.

It is easily possible to work out, to any order, the coupling of the 
hypothetical ``$\Theta$-pion'' $SU(3)$ multiplet to the usual pion 
$SU(3)$ multiplet.  Due to the rigid restrictions imposed by the 
$SU_L(3) \otimes SU_R(3)$ geometry, these couplings are completely 
determined. However, once more in analogy to the electroweak 
paradigm, we would expect there to be a large mass gap between these 
(even if they existed as bound states of sufficient lifetimes to
be detectable) and the usual pion octet that would defeat the naive 
hope of detection.   To lowest order in the $\Theta$-pion field, the 
couplings that follow from (10) take the form
$$
{\cal S}_{{1^{st}}-order \, int.} (\Theta) = - \Big( 
\frac {f_{\pi}^2}{2 C_2} \Big)  \int d^4 x \, \Big\{ ~  
\Theta^i ~ [ \, {\cal T}_{~ i} ~-~ ( \pa^{\un a} {\cal 
V}_{\un a \, i} \, ) \, ]~\Big\} ~~~, 
{~~~~~~~~~~~~~~~~~~~}$$
$$ \eqalign{
{\cal T}_{~p} ~=~  &g_1 {\rm {Tr}} \Big[\, \frac{\pa {~~} }
{\pa \Pi^p } [ \Big( \frac{\pa U\dg }{ \pa \Pi^i } \Big)  \Big( 
\frac {\pa U}{ \pa \P^j} \Big) ] \, \Big] ~ (\, \pa^{\un a} 
\P^i )  (\, \pa_{\un a}  \P^j ) \cr
&+ \, g_1 {\rm {Tr}} \Big[\, \frac{\pa {~~} }
{\pa \Pi^p } [ \Big( \frac{\pa U\dg }{ \pa \Pi^i } \Big)  \Big( 
\frac {\pa^2 U}{ \pa \P^j  \pa \P^k} \Big) \, + \, {\rm {h.\, c.}}
] \, \Big] ~ (\, \pa^{\un a} \P^i )  (\, \pa_{\un a}  \P^j )  \P^k \cr
&- \, g_2 {\rm {Tr}} \Big[\, [ \Big( \frac{\pa^2 U\dg }{\pa \Pi^i 
\pa \Pi^j} \Big)  \Big( \frac {\pa^2 U}{ \pa \P^k  \pa \P^p} \Big) 
\,  \, + \, {\rm {h.\, c.}}] \, \Big] ~ (\, \pa^{\un a} \P^i )\, 
\P^j (\, \pa_{\un a}  \P^k )  \cr
&+ \, g_1 {\rm {Tr}} \Big[\, \frac{\pa {~~} } {\pa \Pi^p } [ \Big( 
\frac{\pa^2 U\dg }{\pa \Pi^i \pa \Pi^j} \Big)  \Big( \frac {\pa^2 
U}{ \pa \P^k  \pa \P^l} \Big) \, ] \, \Big] ~ (\, \pa^{\un a} \P^i 
)\, \P^j (\, \pa_{\un a}  \P^k ) \, \P^l 
~~~, \cr
}$$
$$ \eqalign{
{\cal V}_{\un a \, i} ~=~  &g_2{\rm {Tr}} \Big[\, \Big( \frac{\pa 
U\dg }{ \pa \Pi^i } \Big)  \Big( \frac {\pa^2 U}{ \pa \P^j \pa \P^k} 
\Big) \, + \, {\rm {h.\, c.}} \, \Big] ~  (\, \pa_{\un a}  \P^j )  
\P^k {~~~~~~~~~~~~~~~~~~~~}  \cr
&- \, g_1 {\rm {Tr}} \Big[\,  \Big( \frac{\pa U\dg }{\pa \Pi^j} 
\Big)  \Big( \frac {\pa^2 U}{ \pa \P^i  \pa \P^k} \Big) 
\,  \, + \, {\rm {h.\, c.}} \, \Big] ~ (\, \pa_{\un a}  \P^j ) 
\P^k  \cr
&+ \, g_1 {\rm {Tr}} \Big[\, \Big( \frac{\pa^2 U\dg }{\pa \Pi^i \pa 
\Pi^j} \Big)  \Big( \frac {\pa^2 U}{ \pa \P^k  \pa \P^l} \Big)  
\, + \, {\rm {h.\, c.}} \, \Big] ~ \P^j (\, \pa_{\un a}  \P^k ) \, \P^l ~~~, 
}\eqno(20)  $$
where the constants $g_i$ are defined by
$$
g_1 ~=~ sin^2 (\g_S) \, tan(\g_S) ~~,~~ g_2 ~=~ sin(\g_S ) \, cos(\g_S ) 
~~~, 
\eqno(21)  $$
The derivation of (20) follows from a Taylor expansion w.r.t $\Theta^i$ 
and the use of (18).  We note that $g_1 = O[ (\g_S)^3]$ and $g_2 = O[ 
(\g_S)]$ so that the leading vertex for the emission of the $\Theta$-pion
is obtained from the first term of ${\cal V}_{\un a \, i}$. The next
such vertex is obtained from the third term in ${\cal T}_i$.
We emphasize that the group elements $U$ that appear in (20) are
evaluated in the limit of $\Theta \to 0$.

Having understood this simple mechanism for describing the pion in 
a manner where it is a linear combination of two {\it {distinct}} 
representations of the $SU_L(3) \otimes SU_R(3)$ group manifold, it 
should be natural to ask, if it possible to have even more complicated 
descriptions of the pion.  The answer is yes.  The key point is 
that higher order tensor fields $B^{i_1 ... i_p}$ over the $SU_L(3) 
\otimes SU_R(3)$ group manifold possess infinitesimal variations 
of the form
$$
\d B^{i_1 ... i_p} ~=~  \a^{(A)}_1 \cdots \a^{(A)}_p \Big( \pa  
\xi^{i_1}_{(A)} / \pa A^{j_1} ~ \Big) \cdots  \Big( \pa  
\xi^{i_p}_{(A)} / \pa A^{j_p} ~ \Big) B^{j_1 ... j_p} 
{}~~~. \eqno(22)  $$
Multiplying by factors of $\Big( \pa U / \pa A^i \Big)$ and
$\Big( \pa U\dg / \pa A^i \Big)$ appropriately, such higher
tensor fields can be converted into matrix valued fields whose
actions take precisely the same form as the last term in 
(10).  Clearly linear combinations of such actions lead
to even more complicated descriptions and many more mixing angles.

The observations that we have made in this section amount to
mathematical curiosities. There is no particular motivation for 
choosing any of the ur-formulations of the pion over the 
standard one.  The presence of supersymmetric models with
Dirac-like pionini has the potential to change this 
dramatically.

\section{Chiral Superfield Fibers Over Chiral Superfield Manifolds}

~~~~In the third work of ref. \cite{SJG}, the general $\s$-model
geometry of a CNM model was introduced.  The distinctive feature
that differentiates this from the usual 4D, N = 1 supersymmetric
non-linear $\s$-model geometry is that the nonminimal superfields 
can be introduced as 1-forms {\it {not}} coordinates of the 
$\s$-model manifold.  However, once this geometrical feature is 
realized, it becomes obvious that in a general chiral superfield 
non-linear $\s$-model with coordinates $(\Phi^1 , \, \Phi^2 , ... 
, \Phi^{2p})$, it is always possible to regard a subset (say 
$(\Phi^1, \, \Phi^2 , ... , \Phi^p)$) as the coordinates of a 
sub-manifold and replace the remaining chiral superfields by 
1-forms over the sub-manifold.  For an $SU_L(3) \otimes SU_R(3)$ 
group manifold (which is what we require in the present setting) 
this idea may be implemented in the following way.

It is a well known fact that in order to describe a Dirac
spinor requires two distinct superfields. So if the pionini
existed as Dirac particles, two $SU(3)$ octets of chiral
superfields are required. We may denote these by ${\Phi
}_R^{\rm I}$ and ${\Phi}_L^{\rm I}$ with ${\rm I} = 1,...,8$. 
Let us for the moment assume that both transform as 
the coordinate of the $SU_L(3) \otimes SU_R(3)$ group
manifold.  If this is the case, we may exponentiate 
appropriately either one to form an $SU(3)$ group element. 
For this purpose let us pick ${\Phi}_R^{\rm I}$ so that $U(
{\Phi}_R )$ is such a group element. We may define a new chiral
superfield variable $\chi^{\rm K}$ though the equation ${\Phi
}_L^{\rm I} \equiv {C_2}^{-1} Tr [~ ( \pa_{\rm K} U({\Phi}_R ) 
)\, \chi^{\rm K} t^{\rm I} ]$. As a consequence of this 
definition, $\chi^{\rm I}$ must be a 1-form over the group 
manifold.  Thus, an equivalent set of variables is given by 
${\Phi}_R^{\rm I}$ and $\chi^{\rm I}$. In this way Dirac-like 
pionini leads to the introduction of chiral superfields that 
are distinct representations of the group manifold.

Once one is forced to utilize two chiral superfields, then the
pion might occur as a linear combination of the two psuedo-scalar 
component fields that occur at leading order in the $\q$-expansion.
This leads to the introduction of a mixing angle and thus the 
simplest ur-formulation, as presented in section two, seems to 
arise whenever we demand the presence of Dirac-like pionini.
This is independent of whether we utilize two chiral superfields
or one chiral superfield and one nonminimal superfield 
as in the CNM models.

So we may begin by introducing a set of chiral superfields 
$\Phi^{\rm I}$ that correspond to the coordinates of the $SU(3)$ 
Lie-algebra.  Next we introduce a {\it {distinct}} set of $SU(3)$ 
1-form chiral superfields $\chi^{\rm I}$.  The infinitesimal 
transformation law of $\Phi^{\rm I}$ is now determined by (3) 
replacing all ordinary fields by superfields. Similarly, the 
infinitesimal transformation laws of $\chi^{\rm I}$ are determined 
by (8) where again all ordinary fields are replaced by 
superfields.

Upon defining the superfield group elements $U$  by
$$
{U} (\Phi) ~\equiv~ \exp \Big[ {{ \,{\Phi}^{\rm I} t_{\rm 
I}}\over{~ f_{\pi}\, cos (\g_{\rm S}) ~}} \Big] ~~~~,
\eqno(23) $$
we may take for an action\footnote{For the sake of simplicity, we 
will ignore all other higher derivative terms as well as the the 
issue \newline ${~~\,~~}$ of mass terms.},
$$\eqalign {
{\cal S}(\Phi,\chi) ~=~ \Big( \frac {f_{\pi}^2}{ C_2} \Big) 
cos^{2}(\g_S) &\int d^4 x \, d^2 \q \, d^2 {\bar \q} \Big\{ 
~ \Big( \, {\cal N}_0  \, Tr [ U\dg U ] ~+~  {\cal N}_1 Tr [\, 
\Hat {\chi}\dg \Hat \chi \,] \, \Big) {~~~~~~~~~} \cr
- & \Big[~ i {\cal H}^{1}_{{\rm I} \, {\bar {\rm J}} \, {\rm K} 
\, {\rm L}} (\Phi, \, {\Bar \Phi}) \, \chi^{\rm I} ({\Bar D}^{
\Dot\b} {\Bar \Phi}^{\rm J} ) (D^{\g}\Phi^{\rm K}) (\pa_{\g \Dot
\b}\Phi^{\rm L}) ~+~ {\rm {h.c.}} ~ \Big]  \cr
- & \Big[ i {\cal H}^{2}_{{\bar {\rm I}} \, {\bar {\rm J}} \, 
{\rm K} \, {\rm L}} (\Phi, \, {\Bar \Phi}) \, {\Bar \chi}^{\rm 
I} (  {\Bar D}^{\Dot \b} {\Bar \Phi}^{\rm J}  )   (D^{\g}\Phi^{
\rm K}) (\pa_{\g\Dot\b} \Phi^{\rm L})~+{\rm {h.c.}} ~\Big] 
\Big\} \cr 
&+~{\cal S}_{NR} (\Phi)  ~~~,}
\eqno(24) $$
where $S_{NR}$ refer to the Rohm-Nemanschansky action \cite{E}.  Due 
to the work of these authors, we know that the ultimate term above 
contains a component level WZNW term.  With a little modification 
of their calculations, it is easy to see that third and fourth 
actions above also contain terms that may be interpreted as WZNW terms.  
So for arbitrary values of $\g_S$, this action always contains a 
component level WZNW term.  Finally we note that the action in (24) 
does {\it {not}} correspond to the dual formulation of the CNM models 
of \cite{JOMMM}.  This is most easily verified by noting that only 
one factor of $\chi$ appears in the WZNW terms above.  The dual 
CNM models always are quadratic in this dependence.  By the 
same reasoning, we see that the NR action is also not the dual 
CNM model action.

The $SU_L(3) \otimes SU_R(3)$ symmetry of this action is guaranteed 
as long as ${\cal H}^{1}_{{\rm I} \, {\bar {\rm J}} \, {\rm K} \, 
{\rm L}}$ and ${\cal H}^{2}_{{\bar {\rm I}} \, {\bar {\rm J}} \, 
{\rm K} \, {\rm L}}$ are appropriately transforming tensors, i.e. 
respectively of type (3,1) and (2,2) with respect to a 
holomorphic-anti-holomorphic form basis.

On the other hand, the first two terms in (24) obviously describe 
a particular 4D, N = 1 supersymmetric non-linear $\s$-model with 
a K\" ahler geometry. We now want to focus on the component field 
action described by these terms.  First we define the component 
field via the following definitions (with $\Phi \equiv \Phi^{\rm I} 
t_{\rm I})$,
$$ \eqalign{ {~~~~}
{\Phi} | ~\equiv~ {A}^{\rm I}(x) t_{\rm I} &=~ \Big\{ \, {\cal A}^{\rm 
I} (x)~+~ i \Big[ \, \Pi^{\rm I} (x) cos(\g_{\rm S} ) ~+~ \Theta^{
\rm I} (x) sin(\g_{\rm S} ) \, \Big] \, \Big\} t_{\rm I} ~~~, \cr
\Big( D_{\a} U \Big) | &\equiv~ \psi_{\a} (x)  ~~~~, ~~~~
\Big( D^2 U \Big) | ~\equiv~ {\rm F} (x)  ~~~, 
}\eqno(25) $$
where we observe that upon setting ${\cal A}^{\rm I} = 0$, this reduces 
back to  (7). In a similar manner we define,
$$  \eqalign{
\Hat {\chi} | ~\equiv~ {\Hat B}(x) &=~ [\, \pa_{\rm I} U(\Phi |)  \, ] 
\, \Big\{ ~{\cal B}^{\rm I}(x) ~+~ i \Big[ \,  - \Pi^{\rm I} (x) sin(
\g_{\rm S} ) ~+~ \Theta^{\rm I} (x) cos(\g_{\rm S} ) \, \Big] ~ \Big\}  
~~~, \cr
\Big( D_{\a} \Hat {\chi} \Big) | &\equiv~ \z_{\a} (x) ~~~~, ~~~~
\Big( D^2 \Hat {\chi} \Big) | ~\equiv~ {\rm G} (x)~~~.
}\eqno(26) $$
where $\pa_{\rm I} U(\Phi |) = \Big( \pa U (\Phi |) / \pa A^{\rm I} 
\Big)$. Note that the component denoted by ${\Hat B}$ here is slightly 
different from the similar field defined in (10).  Only upon setting 
${\cal A}^{\rm I} = {\cal B}^{\rm I} = 0$ will the two fields coincide.

After a direct calculation we find,
$$ \eqalign{ {~~}
&\Big( \frac {f_{\pi}^2}{ C_2} \Big) cos^{2}(\g_S) 
\int d^4 x \, d^2 \q \, d^2 {\bar \q} ~ \Big( \, {\cal N}_0 
\, Tr [ U\dg U ] ~+~  {\cal N}_1 Tr [\, \Hat {\chi}\dg \Hat 
\chi \,] \, \Big)  ~= \cr
&{~~~~}\, \Big( \frac {f_{\pi}^2}{ C_2} \Big) {\cal N}_0 cos^{2}(\g_S)
\int d^4 x \, Tr \Big[ \, - \frac 12 [ \pa^{\un a} {U}\dg (\Phi |) 
\,] \, [ \pa_{\un a} {U (\Phi |)} \, ] ~-~ i  {{\bar \psi}^{\ad}}\dg 
\, \pa_{\un a} {\psi}^{\a} ~+~ {\rm F}\dg {\rm F} ~ \Big] \cr
&+~ \Big( \frac {f_{\pi}^2}{ C_2} \Big) {\cal N}_1 cos^{2}(\g_S)
\int d^4 x \, Tr \Big[ \, - \frac 12 (\pa^{\un a} {\Hat B}\dg \, ) (
\pa_{\un a} {\Hat B} \, ) ~-~ i {\z}^{\a}  \pa_{\un a} {{\bar 
\z}^{\ad}}\dg  ~+~ {\rm G}\dg {\rm G} ~ \Big]  ~~~.} 
\eqno(27) $$

Using the same methods as were developed in the discussion of the
$\Theta$-pion multiplet coupling to the usual pion multiplet
in section two, similar results can be derived from (27) for the 
coupling to the ${\cal A}$ and ${\cal B}$ multiplets. As well it 
is straightforward to see that using the definitions above,
the Dirac pionino defined by
$$\ell^{\rm I}(x)  ~\equiv ~ 
\left(\begin{array}{c}
 \psi^{\rm I}_{\a}(x) \\
~\\
{{\Bar \z}{}^{\rm I}_{\ad}}\dg (x)\\
\end{array}\right) ~~~~~~, ~~~~~~~
\begin{array}{c}
 \psi^{\rm I}_{\a}(x) \, =\,  \frac 12 ( {\rm I} + \g^5 ) \, \ell^{
\rm I} (x)  \\
~\\
{{\Bar \z}^{\rm I}_{\ad}}\dg (x) \, =\,  \frac 12 ( {\rm I} - 
\g^5 ) \, \ell^{\rm I} (x) \\ \end{array} ~~~~~~, 
\eqno(28) $$
is a free field (at this level) with interactions only arising through
the WZNW terms in (24).  Thus, the fermionic terms of (27) can be
combined as
$$
i ~ {\Bar \ell}\dslash \, {\ell} ~=~ i \, {{\bar \psi}^{\ad}}\dg \, 
\pa_{\un a}{\psi}^{\a} ~+~ i \, {\z}^{\a}  \pa_{\un a} {{\bar \z}^{
\ad}}\dg ~~~.
\eqno(29) $$
Although the sum\footnote{By further field re-definitions, the spinor 
action in (27) can be made to possess a canonical \newline ${~~~\,~}$ 
normalization.} in (29) is invariant under the exchange of $\psi_{\a}$ 
and ${\bar \z}_{\ad}\dg$, (thus insuring parity invariance) the total 
action in (24) does not possess this symmetry.  Thus, we find the result 
that the mere presence of Dirac-like pionini implies a breaking of 
parity even in the absence of the electroweak interaction. The breaking 
of parity can also be observed in the form of the supersymmetry variations
that leave the action invariant. The left and right components of $\ell$ 
enter these variations in a highly asymmetrical manner. A similar result 
has been observed within the CNM models.  Although the kinetic energy 
terms for the spin-0 fields in (27) appear to be the same as that in 
(10), this is only an apparent similarity. The terms in (27) include 
kinetic energies for ${\cal A}^{\rm I}$ and ${\cal B}^{\rm I}$ which 
do not appear at all in (10). Similarly, the field ${\cal A}^{\rm I}$ 
appears to all powers in (27) while ${\cal B}^{\rm I}$ appears 
only quadratically.  This nonpolynomial appearance of ${\cal 
A}^{\rm I}$ can be understood upon observing that $U\dg (\Phi |)
\ne U (\Phi |)$ which implies that the sigma model in (27)
possesses a non-compact geometry in contrast to the compact
geometry described by the model in (10).  One of the most obvious 
points is that if we only retain the fields $\Pi^i$ and $\Theta^i$ 
in (27) and set all remaining fields to zero, then the superfield 
action in (27) reduces to the action in (10), the pure pion limit 
of which is given by ${\Tilde {\cal S}}_{\s} (\Pi)$ defined in (15).  
By the same duality argument used in \cite{JOMMM}, this must also be 
the pure pion limit of the CNM $\s$-model of the supersymmetric QCD 
effective action (after eliminating all auxiliary fields).

\section{ Conclusion}

~~~~As we saw in the beginning of this work, the ur-formulations of 
the $SU_L(3) \otimes SU_R(3)$ geometry of the pion exist outside of 
the supersymmetric context.  However, without supersymmetry these 
are very ad hoc and unnatural.  In the presence of Dirac pionini 
this is not the case.  As mentioned already the 
simplest ur-formulation also occurs in the context of our previously 
constructed CNM models \cite{SJG,JOMMM}.  The model in (10) is 
determined (up to constants) totally by the $SU_L(3) \otimes SU_R(3)$ 
geometry not supersymmetry. 

This observation permits us to quantify the statement made in ref. 
\cite{JOMMM}, that ``..., the manifestly supersymmetric action should 
come as close as possible to being in agreement with the Gasser-Leutwyler 
parametrization \cite{gasser}.''  The mixing angle $\g_S$ is a measure 
of this closeness.  The smaller the value of this angle, the closer 
the ur-formulation (${\Tilde {\cal S}}_{\s}$) contained in the CNM 
model comes to exact agreement with the usual model of pure pion 
physics (${{\cal S}}_{\s}$). More remarkably, the CNM model is also 
the only one that requires the mixing angle $\g_S$ to be non-zero 
on {\it {theoretical}} grounds. Thus, we find the remarkable 
situation that the parameter $\g_S$ (assuming it is positive) must 
be bounded (by an equation like $0 < \g_S < \frac 1{10}$) from 
above by experimental viability of the model and from below by the 
fact that theoretically it must contain a pion.

As we have seen in a hypothetical world of unbroken supersymmetry,
it is also possible to describe Dirac pionini with the sole use of 
chiral superfields.  However, in the purely chiral superfield 
formulation of such a theory, all possible values of the mixing 
angle $\g_S$ are allowed {\it {theoretically}}.  Thus, we may 
say that the mixing angle $\g_S$ is {\it {essential}} within 
the context of the CNM model but is not otherwise.  The concept 
of {\it {essential}} versus {\it {inessential}} mixing angles 
can also be seen within the standard model.  The weak mixing 
angle ($\theta_W$) is an essential mixing angle because there 
are certain values of $\theta_W$ that are {\it {not}} allowed 
on purely theoretical grounds (i.e. $sin (2 \theta_W ) \neq 0$). 
On the other hand, the Cabibbo and Kobayashi-Maskawa angles are 
examples of inessential mixing angles.

So it looks like there are many {\it {distinct}} ways to 
construct 4D, N = 1 supersymmetric extensions of the chiral 
model that contains the pion and Dirac pionini.  This is true 
at least if we only require superactions that contain a $\s$-model
term as well as a WZNW term.\footnote{It should be noted that
it is only in the case of the CNM models that it has been 
verified that \newline ${~~\,~~}$ all of the terms of the 
Gasser-Leutwyler parametrization can be extended into fully 
supersym \newline ${~~\,~~}$ -metric analogs. So for example, 
it is not known whether there even exist an auxiliary-free  
\newline ${~~\,\,~~}$ description of the `$L_1$ term' using
only chiral superfields.}  We note, however, that the CNM model 
and its dual are the only ones where the transcedental functions
that determine the {\it {entire}} pion sector of the supersymmetric 
QCD effective action via the Gasser-Leutwyler parametrization, are 
{\it {holomorphic}} functions. 

Finally, it is 
interesting that in our hypothetical universe of ur-formulations, 
the ``$\g_S$-suppressed'' pion interactions as well as the appearance 
of the ``$\Theta$-pions'' emerge as signals in the simplest such 
model. These features are necessarily consequences of the 4D, N = 
1 supersymmetric CNM model of the the QCD effective action as well.  
Since supersymmetry (if it applies at all to our world) is badly 
broken, it is unlikely that we could detect the {\it {purely}} pion 
physics `signals' of supersymmetry in Nature...except by the most 
remarkable serendipity.

\noindent
{\bf {Acknowledgment; }} \newline \noindent
${~~~~}$SJG wishes to thank  M. Ro\v cek for conversations
that led to consideration of this problem.  Additional helpful 
comments by Thomas Cohen are gratefully recognized.

$${~~}$$


\begin{thebibliography}{66}

\bibitem{SJG}S.~J.~Gates, Jr., Phys. Lett. {\bf {365B}} (1996) 
132; idem. ``A New Proposed Description of the 4D, N = 1 Supersymmetric
Effective Action for Scalar Multiplets'' to appear in the
proceedings of the Second International Sakharov Conference on
Physics, May 20-25, Moscow, Russia; idem. Univ. of Maryland Preprint,
UMDEPP 96-99, to appear in Nucl. Phys. B.

\bibitem{JOMMM}S.~J.~Gates, Jr., M.~T.~Grisaru, M.~E.~Knutt-Wehlau,
M.\ Ro\v cek and O.~A.~Soloviev, ``N = 1 Supersymmetric Extension of 
the QCD Effective Action,'' Univ.~of Maryland Preprint UMDEPP 97-27,
hep-th/9612196, to appear in Phys. Lett. B.

\bibitem{chipt}J. Donoghue, E. Golowich and B. Holstein, {\it {Dynamics 
of the Standard Model}}, Cambridge Monographs on Particle Physics, 
Nuclear Physics and Cosmology, (Cambridge Univ. Press, 1992), J. D. 
Walecka, {\it {Theoretical Nuclear and Subnuclear Physics}}, 
(Cambridge Univ. Press, 1992); S. Weinberg, {\it {The Quantum
Theory of Fields, Vol. II}}, (Cambridge Univ. Press, 1996).

\bibitem{1}D. Amati, K. Konishi, Y. Meurice, G.C. Rossi and G. 
Veneziano, ``Non-Perturbative Aspects in Supersymmetric Gauge 
Theories,'' Physics Reports {\bf {162}}, {\bf {4}} (1988) 169-248.

\bibitem{2}S. V. Ketov, ``Solitons, Monopoles and Duality: from 
Sine-Gordon to Seiberg-Witten,''  DESY 96--244, ITP-UH-23/96
(ITP, University of Hannover), hep-th/9611209.

\bibitem{WZ}J. Wess and B. Zumino, Nucl. Phys. {\bf {B70}} (1974) 39.

\bibitem{GS}S. J. Gates, Jr. and W. Siegel, Nucl. Phys. {\bf {B187}} 
(1981) 389; S. J. Gates, Jr., M.T. Grisaru, M. Ro\v cek and W. Siegel, 
{\it {Superspace}} Benjamin Cummings, (1983) Reading, MA., pp. 148-157,
199-200; B. B. Deo and S. J. Gates, Jr., Nucl. Phys. {\bf {B254}} 
(1985) 187.

\bibitem{moha}See for example R. N. Mohapatra, {\it {Unification}} {\it
{and}} {\it {Supersymmetry}}, Springer-Verlag, New York (1986); G. Kane,
{\it {Modern}} {\it {Elementary}} {\it {Particle}} {\it {Physics}},
Addison-Wesley Pub. Co., Reading, MA (1993) and references
therein.

\bibitem{GSW}S. L. Glashow, Nucl. Phys.{\bf {22}} (1961) 579, A. Salam 
and J.C. Ward, Phys. Lett. {\bf {13}} (1964) 168; S. Weinberg, Phys. Rev. 
Lett. {\bf {19}} (1967) 1264; A. Salam, in {\it {Elementary}} {\it {
Particle}} {\it {Theory}} (edited by N. Svartholm) Almquist and Forlag, 
Stockholm, 1968.

\bibitem{E}J. Wess and B. Zumino, Phys. Lett. {\bf {37B}} (1971) 95;
S. Novikov, Sov. Math. Dokl. {\bf {24}} (1981) 222; E. Witten,
Nucl. Phys. {\bf {B223}} (1983) 422; T. E. Clark and S. T. Love, 
Phys. Lett. {\bf {138B}} (1984) 289; D. Nemeschansky and R. Rohm,  
Nucl. Phys. {\bf {B249}} (1985) 157.

\bibitem{gasser}J. Gasser and H. Leutwyler, Nucl. Phys. {\bf {B250}}
(1985) 465.


\end{thebibliography}
\end{document}